\documentclass[final,5p,times,fleqn]{elsarticle}

\usepackage{graphicx}
\usepackage{color}
\usepackage{amsmath, amsfonts, amssymb, bm}
\begin{document}

\begin{frontmatter}
\title{Electron-positron pair production in oscillating electric fields\\ with double-pulse structure}
\author{L. F. Granz}
\author{O. Mathiak}
\author{S. Villalba-Ch\'avez}
\author{C. M\"uller}
\address{Institut f\"ur Theoretische Physik I, Heinrich Heine Universit\"at D\"usseldorf, Universit\"atsstr. 1, 40225 D\"usseldorf, Germany}
\date{\today}
\begin{abstract}
Electron-positron pair production from vacuum in a strong electric field oscillating in time is studied. The field is assumed to consist of two consecutive pulses, with variable time delay in between. Pair production probabilities are obtained by numerical solution of the corresponding time-dependent Dirac equation. Considering symmetric field configurations comprising two identical pulses, we show that the pulse distance strongly affects the momentum spectra of produced particles and even the total production probability. Conversely, in a highly asymmetric situation when the field contains low-intensity prepulse and high-intensity main pulse, we identify a range of field parameters where the prepulse despite its weakness can leave visible traces in the particle spectra.
\end{abstract}
\end{frontmatter}

\section{Introduction}
There exist certain classes of electromagnetic fields in which the quantum vacuum can become unstable as electron-positron pair production occurs \cite{Reviews}. The subject has attracted a sustained interest of theoreticians in recent years because corresponding experimental studies are planned at upcoming high-intensity laser facilities, such as the Extreme-Light Infrastructure \cite{ELI}, the Exawatt Center for Extreme Light Studies \cite{XCELS} or the European X-Ray Free-Electron Laser \cite{XFEL}. 

Electric fields oscillating in time can serve as simplified models for laser pulses. In particular, the field resulting from the superposition of two counterpropagating laser pulses, sharing the same frequency, intensity and polarization direction, may be represented by an electric background oscillating in time, provided the characteristic pair formation length is much smaller than the laser wavelength and focusing scale. Pair production in oscillating electric fields was first studied in the 1970s \cite{Brezin, Popov}. Various interaction regimes can be distinguished by the value of the dimensionless parameter $\xi = |e|E_0/(mc\omega)$, with field amplitude $E_0$, field frequency $\omega$, electron charge $e$ and mass $m$, and speed of light $c$. When $\xi\ll 1$, the process probability shows a perturbative power-law scaling with field intensity. Instead, for $\xi\gg 1$ it exhibits a manifestly nonperturbative exponential dependence on $1/E_0$, similarly to the case of constant electric field \cite{Schwinger}. In between these asymptotic domains lies the nonperturbative regime of intermediate coupling strengths $\xi\sim 1$, where analytical treatments of the problem are very difficult.

The only experimental observation of pair production directly from laser fields has been achieved in the late 1990s via the nonlinear Breit-Wheeler process \cite{SLAC}. It relied on the highly relativistic electron beam at the Stanford Linear Accelerator and an optical laser pulse of moderate intensity ($\xi\lesssim 1$). Pair production purely from laser fields has not been observed yet since the requirements in terms of field intensity and/or frequency are beyond present technical capabilities \cite{plane-wave}.
Against this background, theoreticians have recently been exploring various possibilities how to enhance pair production in strong fields and how to bring their predictions closer to the experimental situation. It was shown that the production process is highly sensitive to the precise form of the applied field \cite{DiPiazza, Alkofer, Mocken, Dunne1, Bauke, Grobe1,DiPiazza2} and systematic analyses to find optimal pulse shapes for pair production were performed \cite{optimization1, optimization2, optimization3, optimization4}. In trains of electric pulses, coherent enhancements due to multiple-slit interferences in the time domain were found \cite{Dunne2, slit, combs, modulation, grating}. Particularly strong amplification of pair production is expected to occur in bichromatic fields, consisting of weak high-frequency and strong low-frequency components \cite{Schutzhold,Orthaber,Grobe2,Akal,Otto}. Modifications of the pair production process due to the spatial dependence and magnetic component of laser fields, as opposed to oscillating electric fields, were also studied \cite{Ruf,Alkofer2,Dresden,Grobe3}.

In the present paper, we study pair production by oscillating electric fields with double-pulse structure in the nonperturbative regime of $\xi\sim 1$. The corresponding time-dependent Dirac equation is solved numerically to obtain the production probabilities for given particle momenta. Two complementary scenarios are considered. First, we assume a symmetric field configuration composed of two identical pulses. Here our focus lies on the influence of the time delay between the pulses, which is shown to strongly affect the pair production process. Remarkably, a proper choice of the delay can substantially enhance the total process probability. Second, we examine pair production in an asymmetric field configuration where a rather weak prepulse precedes a strong main pulse. This part of our study is motivated by the fact that in experiment, due to the generation mechanism, high-intensity laser pulses generally come along with a prepulse of much lower intensity. We demonstrate that a sufficiently weak prepulse can hardly affect the total production probability, but still may leave characteristic imprints on the momentum spectra of produced particles.

Our paper is organized as follows. In Sec.~II we briefly outline our theoretical approach to the problem which was derived in detail previously. Our numerical results are presented in Sec.~III. The case of symmetric electric double pulses is exposed in Sec.~III.A, whereas the asymmetric field configuration is analyzed in Sec.~III.B. Concluding remarks are given in Sec.~V. Relativistic units with $\hbar=c=4\pi\epsilon_0=1$ are used throughout unless otherwise stated.

\section{Theoretical framework}
Our goal is to investigate pair production in oscillating electric fields with double-pulse structure. We chose the field to be linearly polarized in $y$-direction. In temporal gauge, such a field $\vec E(t)=-\dot{\vec A}(t)$ can be described by the vector potential
\begin{eqnarray}
\vec{A}(t) = \big[ A_1(t)+A_2(t-T_1-\delta) \big]\,\vec{\rm e}_y\ ,
\label{A}
\end{eqnarray}
where $\delta$ denotes the time delay between the pulses, both of which having sinusoidal time dependence
\begin{eqnarray}
A_j(t) = \frac{m\xi_j}{e}\,\sin(\omega_j t)\,F_j(t)\ .
\end{eqnarray}
The envelope functions have compact support on $[0,T_j]$ with $T_j=(N_j+1)\frac{2\pi}{\omega_j}$ and are defined as
\begin{eqnarray}
F_j(t) = \left\{ \begin{array}{ll}
\sin^2\!\left(\frac{1}{2}\omega_j t\right) &,\ \ 0\le t < \tau_j \\
1 &,\ \ \tau_j \le t < T_j-\tau_j \\
\sin^2\!\left(\frac{1}{2}\omega_j t\right) &,\ \ T_j-\tau_j \le t\le T_j\\
0 &,\ \ \mbox{otherwise} \end{array} \right.
\end{eqnarray}
with turn-on and turn-off increments $\tau_j = \frac{\pi}{\omega_j}$ of half-cycle duration each. In our numerical computations in Sec.~III we will always assume that the pulse frequencies are equal, $\omega_1=\omega_2\equiv \omega$. A generic form of such a potential is depicted in Fig.~1.

\begin{figure}[b]  
\vspace{-0.25cm}
\begin{center}
\includegraphics[width=0.48\textwidth]{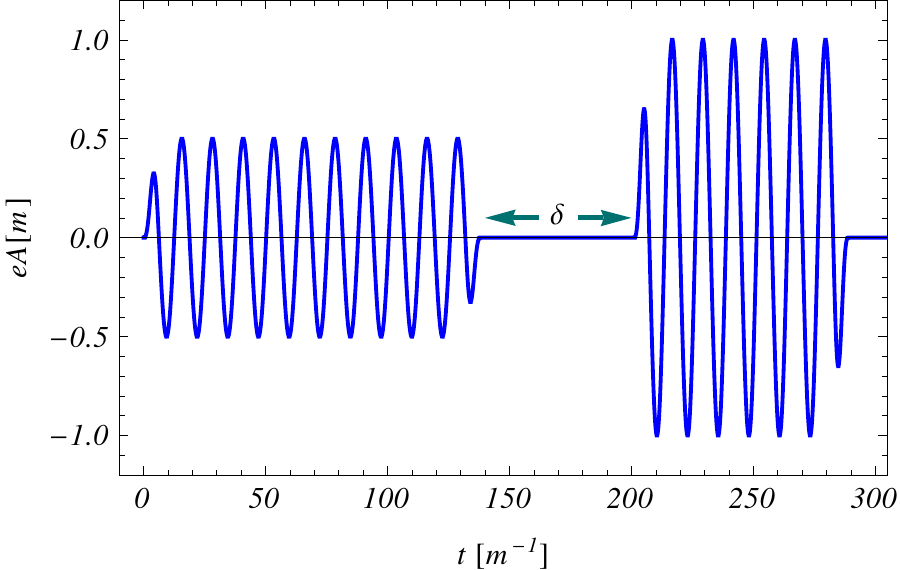}
\end{center}
\vspace{-0.5cm} 
\caption{General form of an oscillating electric field with double-pulse structure. The pulses are characterized by their frequency $\omega_j$, intensity parameter $\xi_j$ and number of plateau cycles $N_j$ ($j\in\{1,2\}$) and have variable time delay $\delta$.}
\label{figure1}
\end{figure}

As shown in \cite{Mocken,Hamlet}, the pair production probability in a time-dependent electric field can be obtained by solving a coupled system of ordinary differential equations which reads 
\begin{eqnarray}
\dot{f}(t) &=& \kappa(t)f(t) + \nu(t)g(t)\ ,\nonumber\\
\dot{g}(t) &=& -\nu^*(t)f(t) + \kappa^*(t)g(t)\ ,
\label{system}
\end{eqnarray}
with 
\begin{eqnarray}
\kappa(t) &=& ieA(t)\,\frac{p_y}{\varepsilon_{\vec{p}}}\ ,\nonumber\\
\nu(t) &=& -ieA(t)\,e^{2i\varepsilon_{\vec{p}} t}\,\left[ \frac{(p_x-ip_y)p_y}{\varepsilon_{\vec{p}}(\varepsilon_{\vec{p}}+m)} + i\, \right]\ .
\label{nu}
\end{eqnarray}
It results from the time-dependent Dirac equation by inserting an ansatz of the form $\psi_{\vec p}(\vec r,t) = f(t)\, \phi_{\vec p}^{(+)}(\vec r,t) + g(t)\, \phi_{\vec p}^{(-)}(\vec r,t)$ where $\phi_{\vec p}^{(\pm)}\sim e^{i(\vec p\cdot\vec r \mp \varepsilon_{\vec{p}} t)}$, with $\varepsilon_{\vec{p}}=\sqrt{{\vec p}^{\,2}+m^2}$, denote free Dirac states with momentum $\vec p$ and positive or negative energy. The suitability of this ansatz relies on the fact that the canonical momentum in a spatially homogeneous external field is conserved, according to Noether's theorem. Outside the time intervall when the field is present, the canonical momentum coincides with the kinetic momentum $\vec p$ of a free particle. Therefore, the invariant subspace spanned by the usual four free Dirac states with momentum $\vec p$ can be treated separately. Due to the rotational symmetry of the problem about the field axis, one may parametrize the momentum vector as $\vec p = (p_x,p_y,0)$ with transversal (longitudinal) component $p_x$ ($p_y$). In addition, there is a conserved spin-like operator, which allows to reduce the effective dimensionality of the problem further.

The time-dependent coefficients $f(t)$ and $g(t)$ describe the occupation amplitudes of a positive-energy and negative-energy state, respectively. The system of differential equations \eqref{system} is solved with the initial conditions $f(0)=0$, $g(0)=1$. After the field has been switched off, $f(T)$ describes the occupation amplitude of an electron state with momentum $\vec p$, positive energy $\varepsilon_{\vec{p}}$ and certain spin projection. Here  $T=T_1+\delta+T_2$ is the total time duration of the double-pulse field. Taking the two possible spin degrees of freedom into account, we obtain the probability for creation of a pair with given momentum as 
\begin{eqnarray}
W(\vec p, T) = 2\,|f(T)|^2\ .
\end{eqnarray}
Note that the created positron has momentum $-\vec p$, so that the total momentum of the pair vanishes.



\section{Results and Discussion}

\begin{figure}[t]  
\vspace{-0.25cm}
\begin{center}
\includegraphics[width=0.48\textwidth]{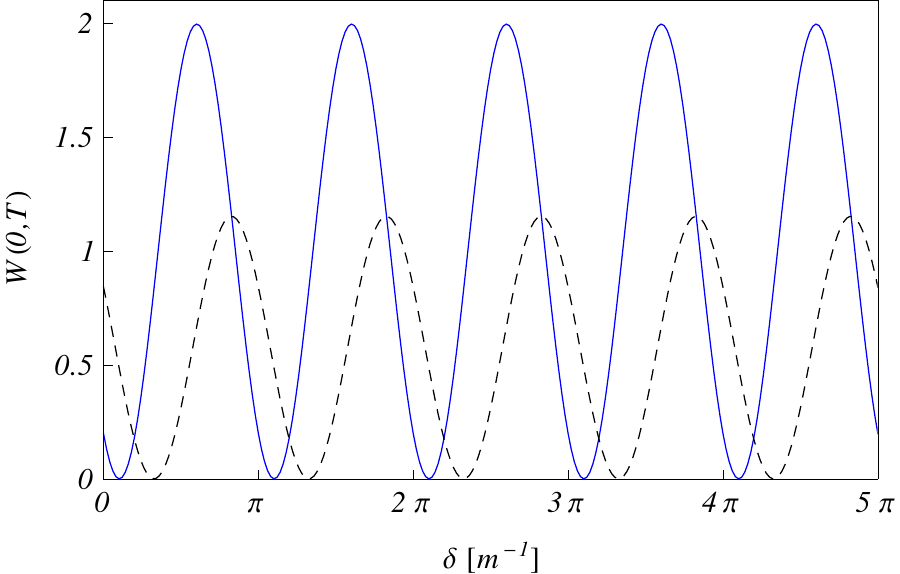}
\end{center}
\vspace{-0.5cm} 
\caption{Pair production probability in electric double pulses as function of the time  delay $\delta$. The field parameters are $\xi_1=\xi_2=1$, $N_1=N_2=6$ with $\omega=0.49072m$ 
(blue solid curve) or $\omega=0.34888m$ (black dashed curve, enhanced by factor 3 for better visibility). The particle momenta are zero, $p_x=p_y=0$.}
\label{figure2}
\end{figure}

We have applied our theoretical approach to calculate field-induced pair production in the nonperturbative coupling regime where $\xi\gtrsim 1$. From previous studies 
based on single electric pulses with periodic time dependence it is known that the pair production shows characteristic resonances whenever the ratio between the energy 
gap and the field frequency attains an integer value. The energy gap amounts to $2\bar{\varepsilon}$ with the time-averaged particle quasi-energy 
\begin{eqnarray}
\bar{\varepsilon}= \frac{1}{T}\int_0^T\sqrt{m^2+p_x^2+[p_y-eA(t)]^2}\,dt\ .
\end{eqnarray}
For example, at $\xi=1$ and $\vec p=0$, one obtains $\bar{\varepsilon}\approx 1.21m$. The enhancement as compared with the corresponding field-free energy [$\varepsilon_{\vec{p}}=m$ 
for $\vec p=0$] is caused by field dressing. Accordingly, frequency values of $\omega=0.49072m$ ($\omega=0.34888m$) were obtained to induce resonant production of particles 
at rest by absorption of five (seven) field quanta (``photons'') \cite{Mocken}. Under these resonant conditions the pair production probability $W(\vec p, T)$ as function of 
the interaction time $T$ exhibits characteristic Rabi oscillations with maximum amplitude of 2.

The frequencies mentioned above have also been applied in our numerical calculations. They were chosen, on the one hand, to highlight the intrinsic phenomenology of pair 
production in electric field double pulses and, on the other hand, to render the computations feasible. Besides, they allow for a direct comparison with the single-pulse
results provided in Ref.~\cite{Mocken}. 

It should  be noted that, at high frequencies ($\omega\gtrsim 0.1m$), significant differences between pair
production in an oscillating electric field and pair production in a standing laser wave arise, due to the spatial dependence and magnetic component of the latter \cite{Ruf,Dresden,Grobe3}. 
Therefore, the numerical results presented below can be transferred to the case of laser fields only in a qualitative manner. The general physical conclusions drawn regarding 
the dependence on the inter-pulse time delay and the potential influence of a prepulse preceding a main pulse, however, are expected to find their counterparts in laser-induced pair
production as well.

\subsection{Symmetric double pulses}

\begin{figure}[t]  
\vspace{-0.25cm}
\begin{center}
\includegraphics[width=0.48\textwidth]{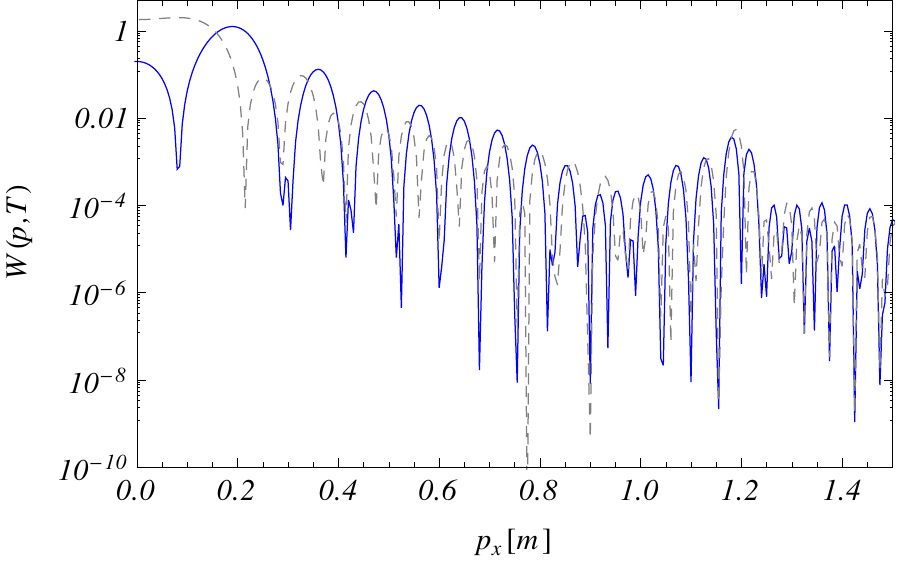}
\end{center}
\vspace{-0.5cm} 
\caption{Transversal momentum distributions of particles created in an electric double pulse with $\xi_1=\xi_2=1$, $\omega=0.49072m$, $N_1=N_2=6$, and time delay $\delta=0$ (blue 
solid curve) or $\delta=\pi/(2m)$ (gray dashed curve). The longitudinal momentum component along the field direction vanishes, $p_y=0$.}
\label{figure4}
\end{figure}

\begin{figure}[t]  
\vspace{-0.25cm}
\begin{center}
\includegraphics[width=0.48\textwidth]{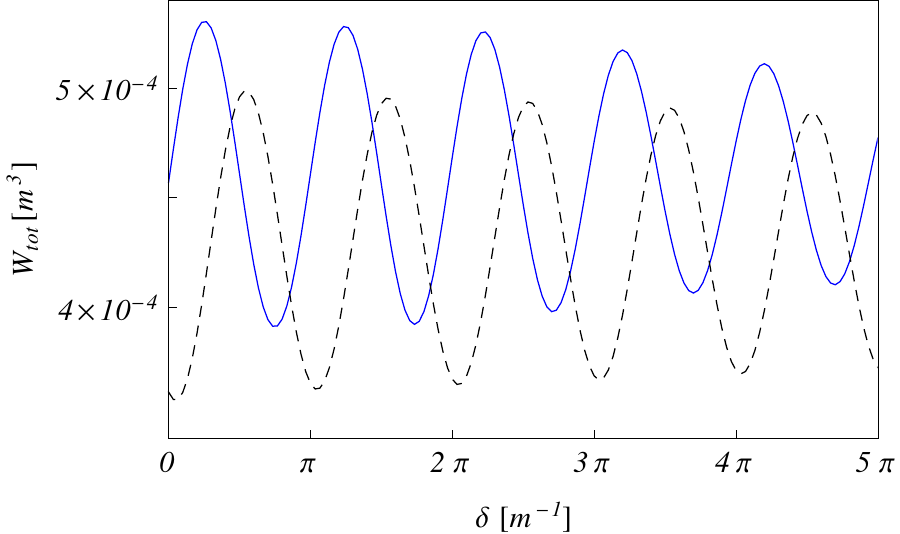}
\end{center}
\vspace{-0.5cm} 
\caption{Total pair production probabilities per Compton volume, as function of the time delay $\delta$. The field parameters are $\xi_1=\xi_2=1$, $N_1=N_2=6$, and $\omega=0.49072m$ 
(blue solid curve) or $\omega=0.34888m$ (black dashed curve, enhanced by factor 15 for better visibility).}
\label{figure5}
\end{figure}

We start our discussion by considering electric fields which are composed of two identical pulses. A key feature of a double-pulse configuration, in contrast to a single-pulse 
field, is the delay time $\delta$. Figure~\ref{figure2} shows the $\delta$-dependence of the probability to produce a particle pair at rest. A characteristic oscillatory behavior 
is found. It arises from the fact that, during the delay period when $A(t)=0$, only the phase factors $e^{\pm i\varepsilon_{\vec{p}} t}$ of the free Dirac states $\phi_{\vec p}^{(\pm)}$ 
evolve in time. The coefficients $f(t)$ and $g(t)$ remain constant instead, as $\dot f(t)=\dot g(t)=0$. Thus, a relative phase factor of $e^{2i\varepsilon_{\vec{p}} \delta}$ develops which 
leads to a periodicity of $\pi/\varepsilon_{\vec{p}}$. In the example of Fig.~2, this value becomes $\pi/m$, in agreement with the data.

In mathematical terms, the initial conditions $f(0)=0$, $g(0)=1$ are transformed by the first pulse to $f(T_1)$, $g(T_1)$. The latter values serve as input for the action of the 
second pulse. During the delay time $\delta$, however, the function $\nu(t)$ has acquired an additional factor of $e^{2i\varepsilon_{\vec{p}}\delta}$ [see Eq.~\eqref{nu}] which introduces 
a relative phase with respect to $\kappa(t)$. Therefore, the action of the second pulse starting at $t=T_1+\delta$ on the ``initial'' values $f(T_1)$, $g(T_1)$ is equivalent to 
the action of the same pulse starting at $t=0$ on the transformed initial values $f(T_1)\,e^{-i\varepsilon_{\vec{p}}\delta}$, $g(T_1)\,e^{i\varepsilon_{\vec{p}}\delta}$. Hence, the mathematical 
effect of the delay time is a change of the ``initial'' conditions for the second pulse by a relative phase. This sort of memory effect can also be seen in connection with 
the well-known  non-Markovian feature of the pair production process \cite{Schmidt:1998vi}.

The phenomenon just described is analogous to Ramsey interferometry in quantum optics. There, a first resonant laser pulse is applied to create a coherent superposition of atomic 
states which, after a variable time delay, is probed by a second laser pulse (``method of separated oscillating fields'') \cite{Ramsey}. During the delay period, the quantum phases 
of the states evolve further, so that the final state population, which is detected after the second pulse has passed, displays a characteristic fringe pattern as function of the 
delay. The underlying coherent quantum dynamics is the same as here. Figure~2 thus represents a Ramsey interferogram of the quantum vacuum. We note that the relation between Ramsey 
interferometry and pair production in two consecutive electric-field pulses has already been established in \cite{Dunne1}. In the present situation this relation is even closer, 
though, since the pulses are sinusoidally oscillating in time (rather than unidirectional) and induce resonant Rabi floppings between the negative- and positive-energy Dirac continua.


Since the delay time enters into the calculation via a complex phase which depends on the particle energy and, thus, the particle momenta, it also affects the momentum distributions 
of created particles. This is illustrated in Fig.~3, which shows the dependence of the pair production probability on the transversal momentum $p_x$ for two different values of $\delta$. 
While the general trend of both curves is similar, including the appearance of a six-photon resonance at $p_x\approx 1.2m$, the pulse delay causes pronounced differences in some regions. 
For example, while many pairs are created with very small momenta $p_x\lesssim 0.1m$ for $\delta=\pi/(2m)$ (gray dashed curve), the production in this region is suppressed for $\delta=0$ 
and rather shifted to somewhat higher momenta $p_x\approx0.2m$ (blue solid curve) where the other curve, in turn, passes through a minimum. Qualitatively similar results are obtained for 
longitudinal momentum distributions (not shown) \cite{Granz}. Thus, by properly chosing the value of $\delta$ one can in principle enhance or suppress the pair production in certain 
regions of momenta.


\begin{figure}[t]  
\vspace{-0.25cm}
\begin{center}
\includegraphics[width=0.48\textwidth]{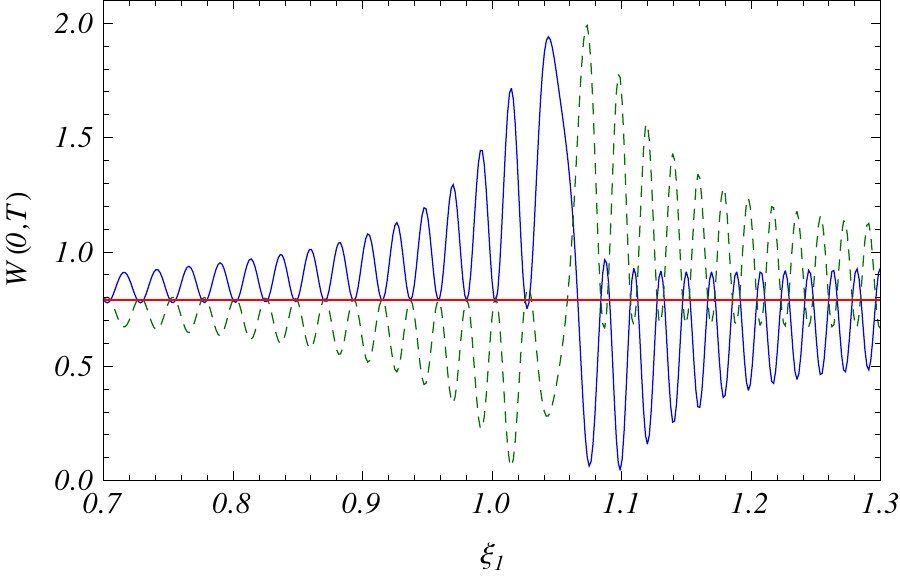}
\end{center}
\vspace{-0.5cm} 
\caption{Pair production probabilities in electric double pulses, as function of the prepulse intensity parameter. The other field parameters are $\xi_2=5$, $\omega=0.5m$, 
$N_1=30$ and $N_2=2$. The time delay is $\delta=0$ (blue solid curve) or $\delta=\pi/(2m)$ (green dashed curve). The particle momenta are zero, $p_x=p_y=0$. 
For comparison, the horizontal red line shows the pair production probability by the main pulse alone.}
\label{figure6}
\end{figure}

By taking an integral over all particle momenta, the total pair production probability (per volume) is obtained. The delay time may have significant impact even on this integrated quantity. 
As depicted in Fig.~4, the total probability shows oscillations with slightly decreasing amplitude when $\delta$ grows. The relative difference between minimum and maximum amounts to about 
25\%. A very similar behaviour of the total probability was observed for pair production by the nonlinear Breit-Wheeler process in two consecutive laser pulses with variable time delay \cite{Jansen} 
(see also \cite{Titov-double} for a related study). For comparison, we note that the total probability in a single pulse comprising 13 cycles amounts to $3.9\times 10^{-4}\; m^3$ ($2.6\times 10^{-5}\; m^3$) for $\omega=0.49072m$ ($\omega=0.34888m$). 
The slow decrease of the oscillation amplitude in Fig.~4 can be understood by noting that the factor $e^{2i\varepsilon_{\vec{p}}\delta}$ becomes more and more sensitive to small variations of $p_x$ 
and $p_y$ when $\delta$ grows. Therefore, the momentum dependence contains increasingly oscillating terms from quantum interference, whose contributions tend to fade out when the integral 
over momenta is taken to obtain the total probability.
 
Our results have demonstrated that a proper choice of the delay time $\delta$ can significantly enhance not only the production of pairs with certain momenta, but also the total production 
probability. This is remarkable because the total electric energy does not change by splitting the field into two parts. The pair production can thus be amplified without ``costs'' in terms 
of the field energy.

\subsection{Asymmetric double pulses}

\begin{figure}[t]  
\vspace{-0.25cm}
\begin{center}
\includegraphics[width=0.48\textwidth]{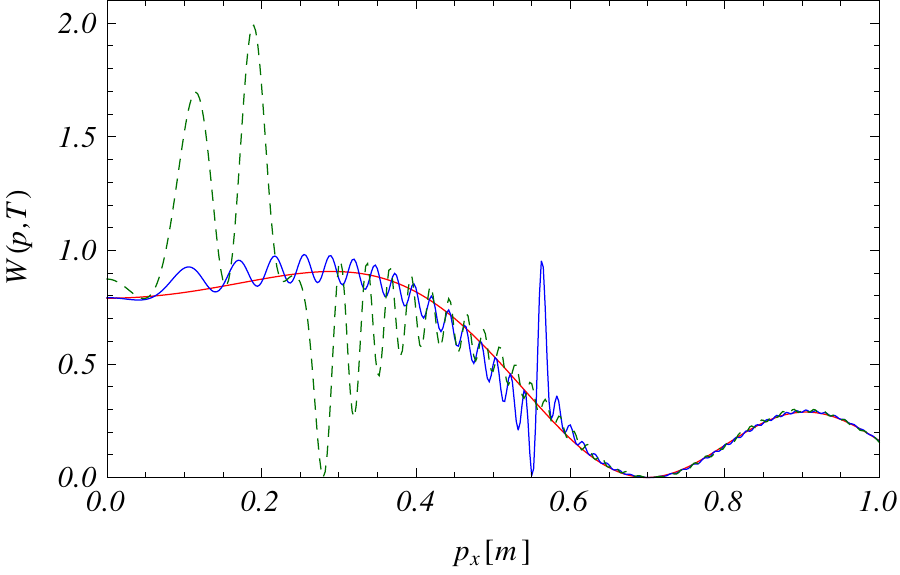}
\end{center}
\vspace{-0.5cm} 
\caption{Transversal momentum distributions of particles created with $p_y=0$ by an electric double pulse with $\xi_2=5$, $\omega=0.5m$, $N_1=30$, $N_2=2$, and $\delta=0$. 
The prepulse intensity parameter is  $\xi_1=0.7$ (blue solid curve), and $\xi_1=1$ (green dashed curve). The solid red line shows the result by 
the main pulse alone.}
\label{figure7}
\end{figure}

Now we turn to pair production in very asymmetric electric double pulses, consisting of a weak multi-cycle prepulse and a strong few-cycle main pulse. The consideration is 
motivated by the fact that, in experiment, a high-intensity short laser pulse is generally accompanied by a much weaker and longer prepulse ($\xi_1\ll\xi_2$, $N_1\gg N_2$). 
Our goal is to reveal for which parameters such a prepulse may have sizeable impact on the pair production process. To put the following discussion into perspective we note 
that a typical contrast ratio in experiment is $\xi_1/\xi_2\sim 10^{-3}$, while prepulse durations of $\sim 1$\,ns have to be compared with $\sim 10$--100\,fs of a main pulse, 
corresponding to $N_1/N_2\sim 10^4$--$10^5$ (see, e.g., \cite{prepulse1,prepulse2}).

\begin{figure}[t]  
\vspace{-0.25cm}
\begin{center}
\includegraphics[width=0.48\textwidth]{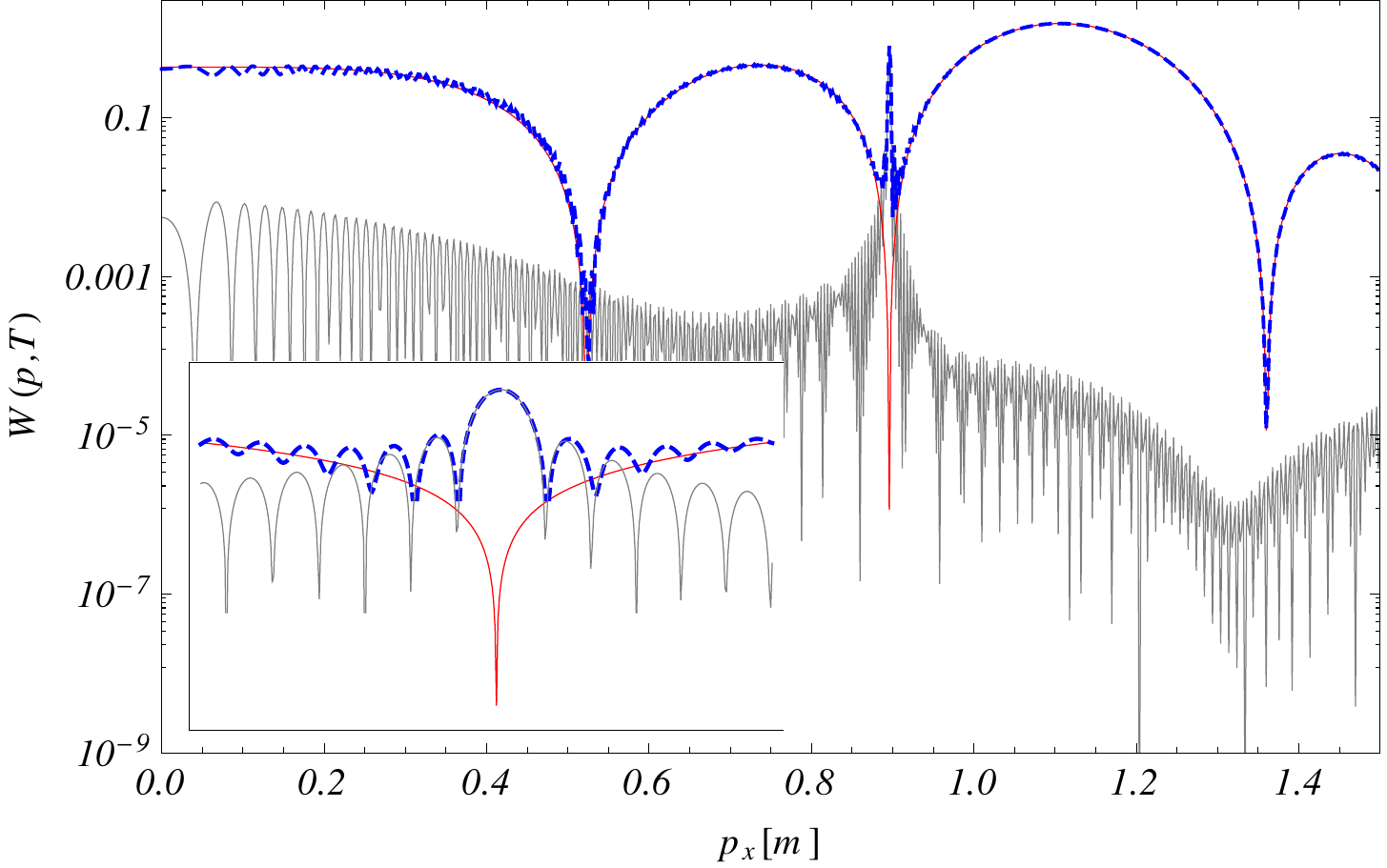}
\end{center}
\vspace{-0.5cm} 
\caption{Transversal momentum distributions of particles created with $p_y=0$ by an electric double pulse with $\xi_1=1$, $\xi_2\approx5.06$, $\omega=0.43342m$, $N_1=91$, $N_2=2$, 
and $\delta=m^{-1}$ (blue dashed curve). The solid red (gray) line shows the result by the main pulse (prepulse) alone. To highlight the peak at $p_x\approx0.9 m$, the inset 
displays an enlargement of the region $0.87m\leqslant p_x\leqslant 0.92 m$.}
\label{figure8}
\end{figure}

Figure~5 shows the probability for producing a pair of particles at rest, as function of the prepulse amplitude. At the given field frequency, the prepulse alone meets the 
condition for a five-photon resonance at $\xi_1\approx 1.05$. Around this point, the production probability is largely modified as compared with the case when the prepulse is 
absent. Apart from an overall oscillatory structure, the probability is generally enhanced for $\xi_1<1.05$ and reduced for $\xi_1>1.05$, provided the main pulse follows immediately 
after the prepulse. For non-zero delay time, the behaviour around the resonance can be inverted with respect to the main pulse result (red line in Fig.~5). Far away from the resonance, 
particularly when $\xi_1\approx 0.7$ falls below $\xi_2$ by an order of magnitude, the influence of the prepulse becomes already quite small, though.

A similar conclusion can be drawn from Fig.~6 which displays a section of the transversal momentum spectrum. The presence of the prepulse is generally visible by oscillations 
which are imprinted on the spectrum arising from the main pulse alone. They can be very substantial when $\xi_1$ is rather large (i.e. $\xi_1/\xi_2\sim 0.2$). However, when 
$\xi_1$ decreases below $\xi_1\lesssim 0.3$ (corresponding to $\xi_1/\xi_2\lesssim 0.05$), these superimposed oscillations die out quickly. In fact, already for $\xi_1=0.3$, 
the pair production probability from a double pulse can hardly be distinguished visually from the corresponding outcome when solely the main pulse is applied (not shown).

Under certain circumstances the influence of a prepulse can be particularly pronounced. Since pair production in an oscillating electric field is a resonant process, the momentum 
distribution of particles produced by the main pulse features both characteristic maxima and minima whose positions depend on the pulse frequency, intensity and duration. It may 
happen that a weak prepulse of long duration produces particles predominantly in a region of momenta where the spectrum from the main pulse alone manifests a deep minimum. Such a 
scenario is illustrated in Fig.~7. Even though the prepulse has only a few percent of the main pulse intensity, it strongly dominates the production of particles with transversal 
momentum around $p_x\approx 0.9m$ and leads to an amplification of the process by two orders of magnitude at this point. Nevertheless, after integrating over all momenta the resulting 
total pair production probability (per volume) remains practically unchanged \cite{Mathiak}.


\section{Conclusion}

Electron-positron pair production from vacuum in oscillating electric double pulses was studied. We have shown that the time delay between two identical pulses strongly affects the 
process by shifting its quantum phases similarly to Ramsey interferometry in atomic physics. A properly chosen delay can selectively modify the momentum spectra of produced pairs 
and significantly enhance even the total production probability, without increasing the applied field energy. It is an intriguing question for future research whether a refined splitting 
of the field into 3 pulses, 4 pulses, and so on could amplify the production probability even further.

Besides, we have demonstrated that a rather weak prepulse preceding a strong main pulse (with $\xi_1/\xi_2\sim 0.1$) can leave visible signatures in the momentum spectra. For contrast 
ratios $\xi_1/\xi_2\lesssim 10^{-3}$, as routinely achieved in high-intensity laser experiments, however, the corresponding influence of a prepulse is expected to be negligible, provided 
the field interacts with perfect vacuum. If, instead, some electrons or atoms are initially present due to imperfect vacuum conditions, a prepulse preceding the main pulse can be relevant 
(see, e.g., \cite{Bell}).

\section*{Acknowledgement}

This study has been performed within project MU 3149/6-1 of the Research Unit FOR 2783 funded by the German Research Foundation (DFG). We thank A.~G\"orlitz for useful conversations on 
Ramsey interferometry and D.~Ising for his help at the onset of this study. L.~F.~G. and O.~M. contributed equally to the present paper.



\begin{thebibliography}{33}
\bibitem{plane-wave} We note that the quantum vacuum remains stable in the presence of a single plane laser wave, no matter how high its frequency or intensity are.
\end{thebibliography}
\end{document}